
\input harvmac
\def\np#1#2#3{Nucl. Phys. B{#1} (#2) #3}
\def\pl#1#2#3{Phys. Lett. {#1}B (#2) #3}

\def\physrev#1#2#3{Phys. Rev. {D#1} (#2) #3}

\def\prep#1#2#3{Phys. Rep. {#1} (#2) #3}

\def\ev#1{\langle#1\rangle}
\def\pf{{\rm Pf ~}}
\def\tilde{\widetilde}

\Title{hep-th/9505006, RU-95-23}
{\vbox{\centerline{Exact Superpotentials, Quantum Vacua
and Duality}
\centerline{in Supersymmetric $SP(N_c)$ Gauge Theories}}}
\bigskip
\centerline{K. Intriligator and P. Pouliot}
\vglue .5cm
\centerline{Department of Physics and Astronomy}
\centerline{Rutgers University}
\centerline{Piscataway, NJ 08855-0849, USA}

\bigskip

\noindent

We study $N=1$ supersymmetric $SP(N_c)$ gauge theories with $N_f$ flavors of
quarks in the fundamental representation.  Depending on $N_f$ and $N_c$,
we find exact, dynamically
generated superpotentials, smooth quantum moduli
spaces of vacua, quantum moduli spaces of vacua with additional massless
composites at strong coupling, confinement without chiral symmetry
breaking, non-trivial fixed points of the renormalization group, and
massless magnetic quarks and gluons.

\Date{4/95}
\lref\ils{K. Intriligator, R.G. Leigh and N. Seiberg, hep-th/9403198,
\physrev{50}{1994}{1092}; K. Intriligator, hep-th/9407106,
\pl{336}{1994}{409}}%
\lref\sv{M.A. Shifman and A. I. Vainshtein, \np{277}{1986}{456};
\np{359}{1991}{571}}
\lref\nonren{N. Seiberg, hep-ph/9309335, \pl{318}{1993}{469}}%
\lref\nati{N. Seiberg, hep-th/9402044, \physrev{49}{1994}{6857}}%

\lref\finnpou{D. Finnell and P. Pouliot, RU-95-14, SLAC-PUB-95-6768,
hep-th/9503115}
\lref\svgc{M.A. Shifman and A. I. Vainshtein, \np{296}{1988}{445};
A. Yu. Morozov, M.A. Olshanetsky and M.A. Shifman,
\np{304}{1988}{291}}
\nref\nonren{N. Seiberg, hep-ph/9309335, \pl{318}{1993}{469}}%
\nref\power{N. Seiberg, The Power of Holomorphy -- Exact Results in 4D
SUSY Field Theories.  To appear in the Proc. of PASCOS 94.
hep-th/9408013, RU-94-64, IASSNS-HEP-94/57}
\nref\ads{I. Affleck, M. Dine and N. Seiberg, \np{241}{1984}{493};
\np{256}{1985}{557}}%
\nref\sv{M.A. Shifman and A. I. Vainshtein, \np{277}{1986}{456};
\np{359}{1991}{571}}
\nref\nsvz{V.A. Novikov, M.A. Shifman, A. I.  Vainshtein and V. I.
Zakharov, \np{223}{1983}{445}; \np{260}{1985}{157};
\np{229}{1983}{381}}%
\nref\cern{D. Amati, K. Konishi, Y. Meurice, G.C. Rossi and G.
Veneziano, \prep{162}{1988}{169} and references therein}%
\lref\om{C. Montonen and D. Olive, \pl {72}{1977}{117}}
\lref\sw{N. Seiberg and E. Witten, hep-th/9407087, \np{426}{1994}{19};
hep-th/9408099, \np{431}{1994}{484}}%
\lref\sem{N. Seiberg, hep-th/9411149 , \np{435}{1995}{129}}%
\lref\isson{K. Intriligator and N. Seiberg, hep-th/9503179, to appear in
Nucl. Phys. B}

Supersymmetric gauge theories have the special property that
certain quantities are holomorphic and can often be obtained exactly
\nonren. See \power\ for a recent review and \refs{\ads - \cern} for earlier
work.  The exact results so obtained provide insight into the dynamics
of strongly coupled gauge theories, revealing a variety of interesting
phenomena.  It is of interest to know which phenomena are specific to
particular examples and which are generic.  Because the particular
details appear to be very specific to the choice of gauge group
and matter content, it is important to consider many different examples.

We analyze
$N=1$ supersymmetric $SP(N_c)$ gauge theory with $N_f$
flavors of
matter in the fundamental representation.
The phenomena which we find are qualitatively
similar to those found in $SU(N_c)$ with matter in the fundamental \refs{
\nati ,\sem} and in $SO(N_c)$ theories with matter in the $N_c$ \refs{\sem
,\isson }.  In terms of the specific details, the $SP(N_c)$ theories
actually provide the simplest examples of these phenomena.
An underlying reason for the qualitative similarity is that
these $N=1$ theories all have dual descriptions \sem, generalizing
the Montonen-Olive, electric-magnetic duality
\om\ of $N=4$
theories \ref\dualnf{H. Osborn, \pl{83}{1979}{321};
A. Sen, hep-th/9402032,
\pl{329}{1994}{217}; C. Vafa and E. Witten, hep-th/9408074,
\np{432}{1994}{3}} and $N=2$ theories \sw\ to $N=1$ supersymmetric
theories.  As with the $N=4$ and $N=2$ duality, the $N=1$ duality
exchanges strong and weak coupling effects.  However, in the $N=1$
duality the original electric description and the dual magnetic
description have different
gauge groups and matter content.  In addition, while the $N=4$
and $N=2$ duality are thought to be exact, $N=1$ duality arises only in
the far infrared.  Nevertheless, duality is extremely powerful.  In
fact, all of the results which we obtain can be understood as
consequences of the duality.
The qualitative similarity of the $SU(N_c)$, $SO(N_c)$ and $SP(N_c)$
theories suggests that
the strong coupling phenomena which have been found and, in particular, the
underlying duality might be generic properties of
$N=1$ supersymmetric gauge theories.  Perhaps they even reveal
some general properties of non-supersymmetric, strongly coupled gauge
theories.

A summary of our results is as follows: For $N_f\leq N_c$ we find
dynamically generated superpotentials, associated with
gaugino condensation for $N_f<N_c$ and instantons for $N_f=N_c$.  For
$N_f>N_c$ massless flavors there is a quantum moduli space of degenerate
but physically inequivalent vacua.  The $SP(N_c)$ gauge symmetry is
broken by the Higgs mechanism in the bulk of this space of vacua.
Classically there are singular submanifolds of vacua where some of the
$SP(N_c)$ $W$ bosons become massless.  In the quantum theory we find
that for $N_f=N_c+1$ this moduli space of vacua is smooth, without
additional massless fields anywhere.  Instantons modify the space of
vacua in this case, removing the classically
singular submanifolds.  For $N_f=N_c+2$ the space of
vacua is the same as the classical one but there are additional
massless mesons rather than $W$ bosons on the singular submanifolds.
At the origin of this space there is confinement without chiral symmetry
breaking.
For $3(N_c+1)>N_f\geq N_c+3$ there is a dual description
in terms of a magnetic
$SP(N_f-N_c-2)$ gauge theory \sem , which we analyze in detail.
For $N_f\leq{3\over 2}(N_c+1)$ the magnetic theory is free in the infrared
while for $3(N_c+1)>N_f>{3\over 2}(N_c+1)$ the theory is in an
interacting $SP(N_f-N_c-2)$ non-Abelian Coulomb phase.
In the interacting range, the fixed point can also be described in terms
of the original electric $SP(N_c)$ theory in a
non-Abelian Coulomb phase.  For $N_f\geq 3(N_c+1)$ the electric theory
is not asymptotically free and is thus free in the infrared.

The unitary symplectic group $SP(N_c)$ is the subgroup of
$SU(2N_c)$ which leaves invariant an antisymmetric tensor $J^{cd
}$, which we can take to be $J={\bf 1}_{N_c}\otimes i\sigma _2$.  The
dimension of this group is $N_c(2N_c+1)$.  We take for our matter content
$2N_f$ fields $Q_i$, $i=1\dots 2N_f$,
in the fundamental $2N_c$ dimensional representation of
$SP(N_c)$.  (As explained in \ref\wittenanom{E. Witten, \pl{117}{1982}{324}},
the number of fundamentals must be even.)
The classical lagrangian has a moduli space of degenerate
SUSY vacua labeled by the expectation values $\ev{Q}$ subject to the
``D-flatness'' constraints.  Up to gauge and global rotations,
the solutions of these
equations are of the form
\eqn\flcqev{Q=\pmatrix{a_1&\ &\ &\ &\ &\ \cr\ &a_2&\ &\ &\ &\
\cr\ &\ &\ . &\ &\ &\ \cr\ &\ &\ &a_{N_f}&\ &\ \cr}\otimes {\bf 1}_2}
for $N_f<N_c$ and for $N_f\geq N_c$ they are of the form
\eqn\fmcqev{Q=\pmatrix{a_1&\ &\ &\ \cr\ &a_2&\ &\ \cr\ &\ &.&\ \cr\ &\
&\ &a_{N_c}\cr\ &\ &\ &\ \cr\ &\ &\ &\ \cr}\otimes {\bf 1}_2}
where, using gauge transformations, the sign of any $a_i$ can be
flipped.  For generic $a_i$ these expectation values break $SP(N_c)$ to
$SP(N_c-N_f)$ by the Higgs mechanism for $N_f<N_c$ and completely break
$SP(N_c)$ for $N_f\geq N_c$.   The space of vacua can be given a gauge
invariant description in terms of the expectation values of the
``meson'' superfield $M_{ij}=Q_{ic}Q_{jd}J^{cd}$.  For
$N_f\leq N_c$, the space of vacua $\ev{Q}$
correspond to arbitrary antisymmetric
expectation values
of the $M_{ij}$; these $N_f(2N_f-1)$ fields are the matter fields
left massless after the Higgs mechanism.  For $N_f> N_c$, the
space of classical vacua $\ev{Q}$
correspond to antisymmetric expectation values
$\ev{M_{ij}}$ with rank$(\ev{M})\leq 2N_c$.  This classical
constraint can be written as
\eqn\classicalcon{\epsilon ^{i_1\dots
i_{2N_f}}M_{i_1i_2}M_{i_3i_4}\cdots M_{i_{2N_c+1}i_{2N_c+2}} =0.}
The $N_f(2N_f-1)$ fields $M$ subject to this constraint are the
light fields in the classical theory, the $4N_fN_c-N_c(2N_c+1)$ fields
left massless after the Higgs mechanism
\foot{There are no
``baryons'' in $SP(N_c)$;
because the invariant tensor
$\epsilon ^{c_1\dots c_{2N_c}}$ breaks
up into sums of products of the $J^{cd}$, baryons break up into
mesons.}.
Note that the space of $\ev{M}$ satisfying the classical constraint
is singular on submanifolds with
rank$(\ev{M})\leq 2(N_c-1)$.  The classical interpretation of this
singularity is that some of the $SP(N_c)$ gauge group is unbroken and
hence there are additional massless $W$ bosons on these submanifolds.

The above classical vacuum degeneracy can be lifted by quantum effects.
In the low energy theory this is represented by a dynamically generated
superpotential for the light meson fields $M$.  The form of a
dynamically generated superpotential is restricted by holomorphy and the
symmetries.  The quantum theory has an anomaly free $SU(2N_f)\times
U(1)_R$ symmetry with the fields $Q$ in the $(2N_f)_{1-{N_c+1\over
N_f}}$ representation.
Holomorphy and these symmetries determine that any
dynamically generated superpotential must be of the form
\eqn\wsymm{W=A_{N_c,N_f}\big({\Lambda _{N_c,N_f}^{3(N_c+1)-N_f}\over \pf
M}\big)^{1/(N_c+1-N_f)},}
where $A_{N_c,N_f}$ are constants and $\Lambda _{N_c,N_f}$ is the
dynamically generated scale.  Because of the constraint \classicalcon ,
this superpotential does not make sense
for $N_f\geq N_c+1$.  Therefore, the vacuum degeneracy is not
lifted for $N_f\geq N_c+1$; for this range of $N_f$ there is a quantum
moduli space of vacua.

Integrating the one loop beta function, the running $SP(N_c)$ coupling
is related to the scale $\Lambda _{N_c,N_f}$ by $e^{-8\pi
^2g^{-2}(E)+i\theta}=(\Lambda _{N_c,N_f}/E)^{3(N_c+1)-N_f}$.  Consider
a theory with $N_f\leq N_c$ in the limit of large $a_{N_f}$ in \flcqev ,
breaking $SP(N_c)$ with $N_f$ flavors to $SP(N_c-1)$ with $N_f-1$ light
flavors at the scale $a_{N_f}$.  We use the $\overline{DR}$ scheme
\finnpou\ in which
the scale $\Lambda _{N_c-1,N_f-1}$ of the low
energy theory is related to that of the high energy theory by matching
the running coupling at the scale set by the masses of the massive charged
$SP(N_c)$ $W$ bosons: $\Lambda _{N_c-1,N_f-1}^{3N_c-(N_f-1)}=2
a_{N_f}^{-2} \Lambda _{N_c,N_f}^{3(N_c+1)-N_f}$.  The low energy theory
has $\pf \widehat M=a_{N_f}^{-2}\pf M$, where $\pf
\widehat M$ runs over the remaining light
fields.  In order for the superpotential
\wsymm\ to properly reproduce that of the low energy theory, the
constants $A_{N_c,N_f}$ must satisfy $A_{N_c,N_f}=2^{N_f/(N_c+1-N_f)}
A_{N_c-N_f,0}$.

Consider starting with the theory with $N_f\leq N_c$ and giving the $N_f$-th
flavor a mass by adding a term $W_{tree}=mM_{2N_f-1,2N_f}$.  The low
energy theory is $SP(N_c)$ with $N_f-1$ flavors with a scale
related to that of the high energy theory by matching the running
coupling at the scale $m$: $\Lambda
_{N_c,N_f-1}^{3(N_c+1)-(N_f-1)}=m\Lambda _{N_c,N_f}^{3(N_c+1)-N_f}$.
Adding $W_{tree}$ to \wsymm\ and integrating out the massive fields,
the superpotential \wsymm\ of the low energy theory is obtained provided
the $A_{N_c,N_f}$ satisfy
$(A_{N_c,N_f}/(N_c+1-N_f))^{N_c+1-N_f}=(A_{N_c,0}/(N_c+1))^{N_c+1}$.
Combined with the condition found in the previous paragraph, this yields
$A_{N_c,N_f}=(N_c+1-N_f)\omega _{N_c+1-N_f}
(2^{N_c-1}A_{1,1})^{1/(N_c+1-N_f)}$, where $\omega _{N_c+1-N_f}$ is an
$N_c+1-N_f$-th root of unity.  As
discussed in \ads , an instanton yields $A_{1,1}\neq 0$; as obtained
in \finnpou , $A_{1,1}=1$ in the $\overline{DR}$ scheme.  The
dynamically generated superpotentials for $N_f\leq N_c$ are therefore
\eqn\lowflavors{W=(N_c+1-N_f)\omega_{N_c+1-N_f} \Big(
{2^{N_c-1}\Lambda_{N_c,N_f}^{3(N_c+1)-N_f}\over
\pf M} \Big)^{1/( N_c+1-N_f)}. }

For $N_f=N_c$ the gauge group is completely broken for $\pf \ev{M}\neq 0$ and
\lowflavors\ is generated by an instanton in the broken $SP(N_c)$.  For
$N_f<N_c$, \lowflavors\ is
associated with gaugino condensation in the unbroken $SP(N_c-N_f)$: as
in \refs{\ads , \svgc, \ils}, $W=(N_c+1-N_f)S_{N_c-N_f}$, where $S=-{1\over
32\pi ^2}W_\alpha ^2$.  The
result \lowflavors\ implies that there is gaugino condensation in
$N=1$ SUSY $SP(N_c)$ Yang-
Mills theory, with ${1\over 32\pi ^2}
\ev{\lambda \lambda}=2\cdot 2^{-2/(N_c+1)}
\omega_{N_c+1} \Lambda^3_{N_c,0}$ (see also \finnpou ).

The superpotential \lowflavors\ reveals that for $N_f\leq N_c$ the
classical vacua are all lifted; the quantum theory has no vacuum at all.
Adding
mass terms $W_{tree}=\half m^{ij}M_{ij}$ to \lowflavors\ gives a
theory with $N_c+1$ supersymmetric vacua
\eqn\mev{\ev{M_{ij}}= \omega_{N_c+1}\left(2^{N_c-1}\pf m\Lambda
_{N_c,N_f}^{3(N_c+1)-N_f}\right)^{1/(N_c+1)}
\left({1\over m}\right)^{-1}_{ij}.}

Consider the theory with $N_f=N_c+1$.  As discussed above, $W_{dyn}=0$ in
this case.  Giving the matter fields masses, their
expectation values must satisfy \mev .  This implies that the
expectation values satisfy
\eqn\qconstr{\pf M=2^{N_c-1}\Lambda _{N_c,N_c+1}^{2(N_c+1)}\qquad
\hbox{for}\qquad N_f=N_c+1.}
Because this is independent of the masses $m^{ij}$, we can take the
masses to zero and the constraint \qconstr\ remains as
a quantum deformation of the
classical constraint \classicalcon\ in the massless theory.  The quantum
correction is due to an instanton.  This is
similar to $SU(N_c)$ with $N_f=N_c$ fundamental flavors \nati.

The classical moduli space of vacua satisfying \classicalcon\ for
$N_f=N_c+1$ had  singular
submanifolds, corresponding to $W$ bosons which classically become
massless.  These singular submanifolds are not on the quantum moduli
space of vacua satisfying \qconstr.  The quantum moduli space of vacua
is smooth; there are no additional massless fields anywhere on the
moduli space.

The $\ev{M}$ expectation values
satisfying \qconstr\ break
the $SU(2N_f)\times U(1)_R$ chiral symmetry to $SP(N_f)\times U(1)_R$.
We should check the 't Hooft anomaly matching conditions
for this unbroken symmetry.  The classical spectrum has fermions in the
$N_c(2N_c+1)\times (1)_1$ and the $2N_c\times (2N_f)_{-1}$.  The
massless spectrum in the quantum theory
are the fluctuations of $M$ around a VEV satisfying
\qconstr ; the fermions are in the $(N_f(2N_f-1)-1)_{-1}$.  The
anomalies do indeed match; with both the classical and the quantum
spectrum we find: $SP(N_f)^2U(1)_R$:
$-2N_cd^{(2)}(2N_f)=-d^{(2)}(N_f(2N_f-1)-1)$, $U(1)_R$: $-N_c(2N_c+3)$,
$U(1)_R^3$: $-N_c(2N_c+3)$, where $d^{(2)}$ is the quadratic
$SP(N_f)$ Casimir.

Turning on a mass for the $N_c+1$-th
flavor, $W=mM_{2N_f-1,2N_f}$ which, integrating out the massive fields
subject to \qconstr , yields the superpotential \lowflavors\ in
the low energy $N_f=N_c$ theory.

We now consider $N_f=N_c+2$.
Giving the matter masses, using \mev , and taking
the masses to zero in various limits, we find that we can explore the
entire moduli space of classical vacua $\ev{M}$ satisfying
\classicalcon.  The classical effective theory has $N_c(2N_c+7)$ light
fields, corresponding to the $M$ satisfying this constraint.
In the quantum theory, the light fields in the low energy theory are all
$N_c(2N_c+7)+6$ possible antisymmetric $M$ with a
superpotential
\eqn\confining{W= -{\pf M\over 2^{N_c-1}\Lambda^{2N_c+1}_{N_c,N_c+2}
}\qquad\hbox{for}\qquad N_f=N_c+2.
} The classical constraint \classicalcon\
arises as the equations of motion of this superpotential.  For
rank$(\ev{M})=2N_c$, \confining\ gives masses to the 6 ``quantum'' components
of $M$ and there are $N_c(2N_c+7)$ fields remaining massless, as
expected classically.
This is similar to $SU(N_c)$ with $N_f=N_c+1$ \nati .

Unlike the $N_f=N_c+1$ case considered above, the vacuum
can be on the singular submanifolds with rank$(\ev{M})\leq 2(N_c-1)$. The
physics of the singularity is that the additional ``quantum''
components of $M$
become massless on these submanifolds.  In particular, at the origin
$\ev{M}=0$, all $(N_c+2)(2N_c+3)$ components of $M$ are massless.  At
the origin the $SU(2N_f)\times U(1)_R$ chiral symmetry is unbroken and
we can check the 't Hooft anomaly matching with this quantum
massless spectrum. They are indeed satisfied; with both the
classical and the quantum
spectrum we find:
$SU(2N_f)^3: 2N_cd^{(3)}(2N_f); SU(2N_f)^2 U(1)_R: -2N_c({N_c+1\over N_c+2}
)d^{(2)}(2N_f);
U(1)_R: -N_c(2N_c+3); U(1)_R^3: {-N_c^3(2N_c+3)\over (N_c+2)^2}$, where
$d^{(2)}$ and $d^{(3)}$ are the quadratic and cubic $SU(2N_f)$ Casimirs.
Note that at $M_{ij}=0$ there is confinement without chiral symmetry
breaking.

Adding
$W_{tree}=mM_{2N_f-1,2N_f}$ to \confining\ and integrating out the
massive fields yields the constraint \qconstr\ in the low
energy $N_f=N_c+1$ theory as an equation of motion.

For $N_f>N_c+2$ massive flavors, a
superpotential which properly reproduces the expectation values \mev\
upon integrating out the massive matter is
\eqn\wlargenf{W=-(N_f-N_c-1)\omega_{N_f-N_c-1} \Big({\pf M \over
2^{N_c-1}\Lambda_{N_c,N_f}^{3(N_c+1)-N_f}} \Big)^{1/
(N_f-N_c-1)}+\half m^{ij}M_{ij}, }
the continuation of \lowflavors\ to these $N_f$.
This superpotential properly describes the theory
only for non-zero $\pf m$.  For $N_f-N_c-1>1$ it has a branch point at $M=0$
indicating additional physics.

We now consider the theories with $N_f>N_c+2$ massless flavors.
As discussed in \sem , the physics at
$\ev{M}=0$ is a non-Abelian Coulomb phase.  For $N_f\geq 3(N_c+1)$ the
theory is not asymptotically free and thus the infrared theory is free.
For $3(N_c+1)>N_f>{3\over 2}(N_c+1)$ there is a non-trivial fixed point
of the renormalization group with massless $SP(N_c)$ glue
and quarks $Q_f$ in an interacting non-Abelian Coulomb phase.  This phase
can be given a dual description in terms of an $SP(N_f-N_c-2)$ gauge
theory with $N_f$ flavors of matter $q^i$
in the fundamental ($i=1\dots 2N_f$),
gauge singlets $M_{ij}$ ($=-M_{ji}$),
and a superpotential\foot{For $SP(1)\cong SU(2)$
there is another dual description, discussed in \sem , in terms of an
$SU(N_f-2)$ gauge theory with $N_f$ flavors.}
\eqn\wdual{W={1\over 4\mu}M_{ij}q^i_cq^j_dJ^{cd}.}
As explained
in \isson , the scale $\Lambda$ of the electric theory,
the scale $\tilde \Lambda$ of the magnetic theory and the
scale $\mu$ in
\wdual\ are related by
\eqn\lltsim{\Lambda _{N_c,N_f}^{3(N_c+1)-N_f}\tilde \Lambda
_{N_f-N_c-2,N_f}^{3(N_f-N_c-1)-N_f}=C(-1)^{N_f-N_c-1}\mu ^{N_f},}
where the constant $C$ will be determined below.  As the electric theory
gets stronger the magnetic theory gets weaker and vice versa.
For $N_f\leq {3\over 2} (N_c+1)$, the electric description of the
Coulomb phase is at very strong coupling and breaks down.  In this range
the magnetic theory is not asymptotically free and is thus free in the
infrared, providing a free description of what looked like very strong
coupling in the electric description.

Taking $M$ to transform as in the electric description, where
$M_{ij}=Q_iQ_j$, the theory \wdual\ has a
global $SU(2N_f)\times U(1)_R$ flavor symmetry with $M$ in the
$(N_f(2N_f-1))_{2(1-{N_c+1\over N_f})}$ and $q$ in the $(\overline{
2N_f})_{N_c+1\over N_f}$.
Note that these symmetries are anomaly free in the dual theory.
At $\ev{M}=0$ the global
$SU(2N_f)\times U(1)_R$ symmetry is unbroken and the 't Hooft anomalies
with the electric spectrum
must match those of the magnetic spectrum, the $SP(N_f-N_c-2)$
glue with $M$ and $q$ matter. The conditions are
indeed satisfied; both spectra give:
$SU(2N_f)^3: 2N_cd^{(3)}(2N_f); SU(2N_f)^2U(1)_R: -2N_c({N_c+1\over
N_f})d^{(2)}(2N_f);
U(1)_R: -N_c(2N_c+3); U(1)_R^3: N_c(2N_c+1) -{4N_c(N_c+1)^3\over N_f^2}$.

The classical equations of motion of
\wdual\ and
the D-terms give $\ev{q^i}=0$, $\ev{M_{ij}}$ arbitrary.
We will see that the classical
condition of the electric theory
that rank$(\ev{M})\leq 2N_c$, which is a classical consequence of
$M_{ij}=Q_iQ_j$ in the electric
theory, will be
recovered by strong coupling effects in the dual theory.
Consider, for example,
giving $M_{2N_f-1,2N_f}$ a large expectation value.  The
corresponding large expectation values of
$Q_{2N_f-1}$ and $Q_{2N_f}$ in \fmcqev\ break the electric $SP(N_c)$
theory with $N_f$ flavors down to a low energy $SP(N_c-1)$ theory with
$N_f-1$ flavors and scale $\Lambda
_{N_c-1,N_f-1}^{3N_c-(N_f-1)}=2M_{2N_f-1,2N_f}^{-1}\Lambda
_{N_c,N_f}^{3(N_c+1)-N_f}$.  The low energy electric theory is weaker.
In the dual theory the large $\ev{M_{2N_f-1,2N_f}}$ gives a large mass
to $q^{2N_f-1}$ and $q^{2N_f}$.  The low energy dual theory is an
$SP(N_f-N_c-2)$ theory with $N_f-1$ flavors of $q^i$, scale $\tilde
\Lambda
_{N_f-N_c-2,N_f-1}^{3(N_f-N_c-1)-(N_f-1)}=(2\mu)^{-1}M_{2N_f-1,2N_f}
\tilde\Lambda _{N_f-N_c-2,N_f}^{3(N_f-N_c-1)-N_f}$ and the
superpotential \wdual .  The low energy dual theory is at stronger
coupling and is the dual of the low energy electric theory.  The
relation \lltsim\ is preserved in the low energy theory.

Giving an expectation value to rank$(\ev{M})$ eigenvalues of $M$, the low
energy dual theory is an $SP(N_f-N_c-2)$ theory with
$N_f-\half$rank$(\ev{M})$ flavors.
We have seen above that $SP(N_c)$ with $N_f\geq N_c+2$ has a vacuum at
the origin, $\ev{Q}=0$, while $SP(N_c)$ with $N_f\leq N_c+1$ does not.
Similarly, in the dual theory, for $N_f-\half$rank$(\ev{M}
)\leq N_f-N_c-1$,
there is no vacuum at $\ev{q}=0$ due to strong coupling effects.
Because the $M$ equation of motion requires the vacuum to be at $\ev{q}=0$,
there is no supersymmetric vacuum for rank$(\ev{M})\geq 2(N_c+1)$.  This
constraint on $\ev{M}$ was an obvious classical constraint in the
electric theory.  In the magnetic theory it is recovered from strong
coupling dynamics and the equations of motion.

Consider giving the $N_f$-th flavor a mass.  The low energy electric
theory is $SP(N_c)$ with $N_f-1$ flavors and scale $\Lambda
_{N_c,N_f-1}^{3(N_c+1)-(N_f-1)}=m\Lambda _{N_c,N_f}^{3(N_c+1)-N_f}$.
This low energy theory is stronger for large $m$.
Adding
$W_{tree}=mM_{2N_f-1,2N_f}$ in the dual theory \wdual, the
$M_{2N_f-1,2N_f}$ equations of
motion give $\ev{q^{2N_f-1}q^{2N_f}}=-2\mu m$, which breaks the dual
$SP(N_f-N_c-2)$ theory to $SP(N_f-N_c-3)$.  Using the $M_{\hat i,
2N_f-1}$ and the $M_{\hat i, 2N_f}$ equations of motion, where $\hat
i=1\dots 2(N_f-1)$, $\ev{q^{\hat i}q^{2N_f-1}}=\ev{q^{\hat
i}q^{2N_f}}=0$.
Using
also the $q^{2N_f-1}$ and the $q^{2N_f}$ equations of motion, the
light fields in the low energy $SP(N_f-N_c-3)$
theory are the singlets $\widehat M_{\hat i, \hat j}$ and the $N_f-1$
flavors of fundamentals $\widehat
q^{\hat i}$ with a superpotential of the form
\wdual .  Using our matching relations for the Higgs mechanism
in the dual theory, the scale of the low energy dual theory is related
to that of the high energy theory by $\tilde \Lambda
_{N_f-N_c-3,N_f-1}^{3(N_f-N_c-2)-(N_f-1)}=-(m\mu )^{-1}\tilde \Lambda
_{N_f-N_c-2,N_f}^{3(N_f-N_c-1)-N_f}$.
The low energy dual theory is weaker for
large $m$.  The relation \lltsim\ is
preserved in the low energy theory.  The low energy dual theory
is the dual of low energy electric theory.

Turning on masses $W_{tree}=\half m^{ij}M_{ij}$ with rank$(m^{ij})=2r$
gives a low energy electric
$SP(N_c)$ theory with $N_f-r$ flavors.  The low energy
magnetic theory is an $SP(N_f-r-N_c-2)$ gauge theory
with $N_f-r$ flavors.  For $N_f-
r-N_c-2=0$, the magnetic gauge group is completely
broken and there is a contribution to the superpotential from instantons
in the broken magnetic gauge group.  This occurs when the low energy
electric theory has $N_f=N_c+2$. Consider starting
{}from the electric theory with
$N_f=N_c+3$ and turning
on $W_{tree}=mM_{2N_f-1,2N_f}$ to go to the low energy theory with
$\widehat N_f=N_c+2$.  In the dual theory the addition of $W_{tree}$ gives
$\ev{q^{2N_f-1}q^{2N_f}}=-2 \mu m$ and, giving masses to the remaining
$q^{\hat i}$ with the $\ev{\widehat M^{\hat i\hat j}}$
of the low energy theory ($\hat i=1\dots 2(N_c+2)$),
an instanton in the broken magnetic $SP(1)$
gauge group produces a low energy superpotential
\eqn\wdualinst{W={\tilde \Lambda _{1,N_c+3}^{6-(N_c+3)}\pf
({1\over 2\mu}\widehat M)\over q^{2N_f-1}q^{2N_f}}=-{\tilde \Lambda
_{1,N_c+3}^{6-(N_c+3)}\pf \widehat M\over (2\mu )^{N_c+3}m}.}
The superpotential \wdualinst\ is
precisely the superpotential in \confining\ provided the constant
appearing in \lltsim\ is given by $C=16$.  We can thus obtain all of our
results for $N_f\leq N_c+2$ from the dual magnetic description of the
theories with $N_f\geq N_c+3$ by flowing down with added mass terms.

Another way to analyze the theory with added mass terms is to consider
the massless theory for generic values of $M$.  The dual quarks acquire
a mass ${1\over 2\mu}M$ and the low energy magnetic theory is a pure
$SP(N_f-N_c-2)$ Yang-Mills theory with scale $\tilde \Lambda
_L^{3(N_f-N_c-1)}=(2\mu )^{-N_f}\tilde \Lambda
_{N_f-N_c-2,N_f}^{3(N_f-N_c-1)-N_f}\pf M$.  Gluino condensation in this
theory leads to
\eqn\dualgcw{\eqalign{W&=(N_f-N_c-1)2\cdot 2^{-2/(N_f-N_c-1)}\tilde
\Lambda _L^3\cr
&=(N_c+1-N_f)\omega_{N_c+1-N_f}
\left({2^{N_c-1}\Lambda _{N_c,N_f}^{3(N_c+1)-N_f}\over \pf
M}\right)^{1/(N_c+1-N_f)},}}
where we used the previously found expression for gaugino condensation
and \lltsim\ with $C=16$.  Adding the mass terms,
this superpotential is the
superpotential \wlargenf\ which, upon adding masses,
yields the correct $\ev{M}$ in \mev\
by its equations of motion.

We have seen that $SP(N_c)$ theories with matter in the fundamental exhibit a
variety of interesting phenomena which appear to be rather generic. In
particular, we have provided evidence for the $N=1$ dual magnetic description
\sem\ which exchanges strong coupling in one theory with weak coupling
in another.  All of our
results can be obtained from this duality.
A deeper understanding of
duality remains to be found; we
hope that these examples will be useful in this regard.

\centerline{{\bf Acknowledgments}}

We would like to thank N. Seiberg for useful discussions and for helpful
comments on the presentation.  This
work was supported in part by DOE grant \#DE-FG05-90ER40559
and by a Canadian 1967 Science fellowship.

\listrefs
\bye